\documentclass{article}

\usepackage{arxiv}

\usepackage[utf8]{inputenc} 
\usepackage[T1]{fontenc}    
\usepackage{hyperref}       
\usepackage{url}            
\usepackage{booktabs}       
\usepackage{amsfonts}       
\usepackage{nicefrac}       
\usepackage{microtype}      
\usepackage{lipsum}		
\usepackage{graphicx}
\usepackage{natbib}
\usepackage{doi}

\usepackage{amsmath}
\usepackage{mathrsfs}
\usepackage{adjustbox}

\newcommand{\colheader}[1]{%
    \makebox[\linewidth][c]{\rule[-0.8ex]{0pt}{2.6ex}\textit{#1}}\\[4pt]
}

\title{An improved periodic activation for PINNs reconstructing convective flows}

\author{
  \href{https://orcid.org/0000-0002-7817-3388}{\includegraphics[scale=0.06]{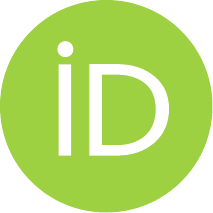}\hspace{1mm}Michael Mommert}$^{1}$ \\
  \And
  \href{https://orcid.org/0009-0003-8963-2724}{\includegraphics[scale=0.06]{orcid.pdf}\hspace{1mm}Marie-Christine Volk}$^{1,2}$ \\
  \And
  \href{https://orcid.org/0000-0003-1838-6194}{\includegraphics[scale=0.06]{orcid.pdf}\hspace{1mm}Christian Bauer}$^{1}$ \\
}

\date{
  \vspace{-2em}%
  \small
  $^{1}$German Aerospace Center (DLR), Institute of Aerodynamics and Flow Technology,
Department Ground Vehicles, Bunsenstr. 10, 37073 Göttingen, Germany \\
  $^{2}$Technische Universität Ilmenau, Institute of Thermodynamics and Fluid
Mechanics, 98693 Ilmenau, Germany
}

\renewcommand{\shorttitle}{Improved periodic activation for PINNs}

\hypersetup{
pdftitle={An improved periodic activation for PINNs reconstructing convective flows},
pdfsubject={},
pdfauthor={M. Mommert et al.},
pdfkeywords={PINN, periodic activation, convective flows},
}

\begin{document}
\maketitle

\begin{abstract}
Architectures with periodic activation functions have already been shown to be beneficial in comparison to monotonic counterparts for a wide range of applications of physics-informed neural networks.
Here, we investigate a network architecture which uses the complex exponential function, generating pairs of sine and cosine outputs as activation functions.
Testing it against comparable, sine-activated multi-layer perceptrons for the task of temperature reconstruction from sparse velocity data for cubic Rayleigh-Bénard convection reveals significant improvements in the reconstruction quality without a substantial increase in computational cost per training step. Vice versa, the improved architecture enables reaching similar reconstruction qualities for a fraction of the expense.
Analyzing the mathematical structure of these networks points to the improvements being rooted in the property of passing both a sine and cosine function forward. 
This way, the subsequent layer is able to adapt the phase of the provided latent periodic functions, and doing it individually for each of its neurons.
\end{abstract}


\section{Introduction}

Periodic activation functions received a prominent introduction by \citet{Sitzmann2020a} for general neural representation applications. Since then, it found its application in physics-informed neural networks (PINNs), which refers to the concept of \citet{Raissi2019}.
These applications comprise the work of \citet{Angriman2023}, who use sine activation functions for PINNs, aiming to fulfill provided higher order statistics through assimilation. 
Further, \citet{Aghaee2024} show that sine activated PINNs yield better results than ones activated by hyperbolic tangent or swish for reconstructing the flow inside blood vessels from sparse data.
A similar comparison of activation functions for the use case of temperature reconstruction in buoyancy driven flows was conducted by \citet{Mommert2024}. In this comparison, sine activation functions performed better than exponential linear unit (ELU) or hyperbolic tangent functions. 
Due to these results, subsequent studies, which use PINNs for three-dimensional reconstruction tasks of planar data \citep{Volk2025} and the application to actual measurement data \citep{Barta2025} in Rayleigh-Bénard convection and turbulent pipe flow \citep{Barta2026}, are all based on periodically activated neural networks.
More recently, \citet{Khodakarami2026} showed that sine activated PINNs are superior to hyperbolic tangent ones for solving the Korteweg–De Vries equation with various optimizers.

A similar approach to introduce periodicity into neural networks are Fourier features \citep{Tancik2020, Wang2021a}. They have an even wider spread into applications and are characterized by carrying both a sine and a cosine function for the same arguments. Yet, Fourier features comprise only the embedding action of an initial network layer. In addition, the frequencies or wavenumbers are not represented by the trainable weights of the layer, but typically determined once by initialization and therefore not subject to training.

The here investigated improved periodic activation represents a synthesis of the two concepts: periodic activation throughout the network and combined sine-cosine outputs. The latter is typical for complex exponential activations, which are already in use for other applications of neural networks \citep{Savchenko2017, Zhang2022}. So, while this concept is not new, the benefits and the reason behind it are typically not discussed. 
We aim to fill this gap by presenting a PINN application for the inverse problem of temperature reconstruction from sparse velocity data, which benefits from a complex exponential activation, and discuss the underlying working principle.

\section{Architecture Description}\label{sec:arch}

To investigate the above-mentioned concept of improved periodic activation, a multi-layer perceptron (MLP) architecture is used. 
Because a paired sine-cosine output is easily expressible in complex numbers, the investigated architecture operates with complex data types.
Please note that, for reasons of layer compatibility within the framework, the complex numbers and operations are still contained within custom layer classes in the practical implementation. 
This means that the layers of the perceptron have two-column inputs and outputs of standard floating point data, which are constructed into or decomposed from complex data within the layer.
For the sake of brevity, we do not cover the changes of data types within the following description.

One single layer $L$ can be described as the following function $\boldsymbol{g}^{(L)}(\boldsymbol{x}^{(L-1)})$ of the complex outputs $\boldsymbol{x}^{(L-1)} \in \mathbb{C}^{N_\mathrm{N}^{(L-1)} \times 1}$ of the previous layer $L-1$, for which $N_\mathrm{N}^{(L-1)}$ is the number of its neurons:

\begin{align}
\boldsymbol{g}^{(L)}(\boldsymbol{x}^{(L-1)}) = \boldsymbol{x}^{(L)} &= \exp \left( i\,\Im \left( \boldsymbol{W}^{(L)}\,\boldsymbol{x}^{(L-1)} + \boldsymbol{b}^{(L)} \right) \right)\label{eq:cexp_layer}
\end{align}

In detail, this layer conducts a matrix multiplication of its inputs with the weight matrix $\boldsymbol{W}^{(L)} \in \mathbb{C}^{N_\mathrm{N}^{(L)} \times N_\mathrm{N}^{(L-1)}}$ and adds a bias vector $\boldsymbol{b}^{(L)} \in \mathbb{C}^{N_\mathrm{N}^{(L)} \times 1}$ to the result. As exponential activations are prone to exploding or vanishing gradients during training, tight restrictions would certainly need to be imposed on the minimum and maximum values of the real part $\Re(\cdot)$ of the linear operation. Therefore, only the imaginary part $\Im(\cdot)$, which produces the periodic outputs, is considered for the exponential activation.

Equal to real numbered MLPs, $N_\mathrm{L}$ layers can be stacked to represent the neural network $\boldsymbol{f}(\boldsymbol{x}^{(0)})$. Regarding the final complex layer, only the real part of its output is used as it is sufficient to describe the fields of fluid flows. As it is typical for our PINN framework, the final layer is a linear combination only \citep{Volk2025}, with real weights $\tilde{\boldsymbol{W}}^{(L)} \in \mathbb{C}^{N_\mathrm{N}^{(L)} \times N_\mathrm{N}^{(L-1)}}$ and biases $\tilde{\boldsymbol{b}}^{(L)} \in \mathbb{R}^{N_\mathrm{N}^{(L)} \times 1}$:

\begin{align}
\boldsymbol{f}(\boldsymbol{x}^{(0)}) &= \boldsymbol{\tilde{W}}^{(N_\mathrm{L})}\,\Re \left( \boldsymbol{g}^{(N_\mathrm{L}-1)} \left(  \boldsymbol{g}^{(N_\mathrm{L}-2)} \left(\ \ldots\ \boldsymbol{g}^{(2)} \left( \boldsymbol{g}^{(1)} \left( \boldsymbol{x}^{(0)} \right) \right) \right) \right) \right)+\boldsymbol{\tilde{b}}^{(N_\mathrm{L})}\label{eq:cexp_net}
\end{align}

The input vectors of a PINN are typically the coordinates of the investigated system in time and space $\boldsymbol{x}^{(0)}=[t,x,y,z]$. Since these are real numbers $\Im(\boldsymbol{x}^{(0)})=\boldsymbol{0}$ applies.

In terms of initialization of the trainable parameters, the following scheme was pursued: All biases as well as the real parts of the weight matrices are set to zero. The only exception are the imaginary parts of the weights, for which a Glorot initialization \citep{Glorot2010} is used. The impact of expanding this initialization type to the real parts is investigated in appendix~\ref{sec:init}.
However, layer $L=1$ represents a a exception as it uses an adapted Glorot initialization \citep{Sitzmann2020a, Mommert2024}, which randomly samples from a uniform distribution $\mathcal{U}\left(-w\sqrt{\frac{6}{N_\mathrm{N}^{(L-1)}}}, w\sqrt{\frac{6}{N_\mathrm{N}^{(L-1)}}}\right)$. The factor $w$ is used to widen the distribution in order to invoke a wider range of basic frequencies or wavenumber to build outputs from. Here, it is set to $w=3$ based on empirical knowledge. 

\section{Impact on test case}
\label{sec:impact}

The following subsections introduce the test case, based on a data set obtained from direct numerical simulation (DNS) of Rayleigh-Bénard convection, as well as details about the PINNs used for the comparison. Subsequently, we present the results for the improved periodic activations in comparison to sine-activated MLP equivalents.

\subsection{Test case description}

The test case is a cubic Rayleigh-Bénard convection flow described by the dimensionless numbers Rayleigh number $\mathrm{Ra}=\hat{g}\hat{\alpha}(\hat{T}_\mathrm{H}-\hat{T}_\mathrm{C})\hat{H}^3/(\hat{\nu}\hat{\kappa})=10^7$ and Prandtl number $\mathrm{Pr}=\hat{\nu}/\hat{\kappa}=0.7$. 
These numbers are defined by the following dimensional (indicated by $\hat{\cdot}$) quantities: the gravitational acceleration $\hat{g}$, the thermal expansion coefficient $\hat{\alpha}$, the thermal diffusivity $\hat{\kappa}$, the kinematic viscosity $\hat{\nu}$, the temperature of the warm bottom surface $\hat{T}_\mathrm{H}$ and the cold top surface $\hat{T}_\mathrm{C}$, as well as the height of the domain $\hat{H}$.
This setup was simulated by fourth-order finite volume DNS in a dimensionless setting (for details see \citep{Volk2025}), with temperatures ranging from $T_\mathrm{C}=-0.5$ to $T_\mathrm{H}=0.5$ and length and time related units referring to the cell height $H=1$ and the free-fall time $t_\mathrm{ff}=\sqrt{\hat{H}/\hat{g}\hat{\alpha}(\hat{T}_\mathrm{H}-\hat{T}_\mathrm{C})}$.
The simulations were governed by the following set of partial differential equations (PDEs) using the Oberbeck-Boussinesq approximation, which limits the effect of temperature changes to a buoyancy force:

\begin{align}
	\partial_t \boldsymbol{u} + (\boldsymbol{u}\cdot\nabla)\boldsymbol{u} &= -\nabla p + \sqrt{\mathrm{Pr}/\mathrm{Ra}}\,\nabla^2\boldsymbol{u} + T \boldsymbol{e}_z \\
    \partial_t T + (\boldsymbol{u}\cdot\nabla) T &= \sqrt{1/(\mathrm{Pr}\,\mathrm{Ra})}\,\nabla^2 T\\
    \nabla\cdot\boldsymbol{u} &= 0
\end{align} 

For our test purposes, we used a data set covering $4.88\,t_\mathrm{ff}$ of the flow within the cubic domain. The objective of the test is to reconstruct the temperature field ($T$) from sparsely provided data of the velocity vector field ($\boldsymbol{u}$). As a side product, the pressure field ($p$) also needs to be reconstructed by the PINN.  

The provided velocity data is sampled in time steps of $0.2\,t_\mathrm{ff}$, which roughly equals the Kolmogorov time scale of this flow.
Spatially, the data is interpolated on locations set by three-dimensional Poisson disk sampling \citep{Cook1986} with a minimal distance of $0.09\,H$, which roughly amounts to ten times the Kolmogorov length scale. These locations do not vary over the time step to avoid a leakage of spatial information throughout time.

\subsection{PINN details}

The PINNs for this reconstruction task use either the architecture described in section~\ref{sec:arch} or a sine-activated MLP ($\tilde{\boldsymbol{f}}$) for comparison.
Analogous to equations~\ref{eq:cexp_layer} and \ref{eq:cexp_net} the latter is defined as follows:

\begin{align}
\tilde{\boldsymbol{g}}^{(L)}(\tilde{\boldsymbol{x}}^{(L-1)})  &= \sin  \left( \tilde{\boldsymbol{W}}^{(L)}\,\tilde{\boldsymbol{x}}^{(L-1)} + \tilde{\boldsymbol{b}}^{(L)} \right)\\
\tilde{\boldsymbol{f}}(\boldsymbol{x}^{(0)}) &= \boldsymbol{\tilde{W}}^{(N_\mathrm{L})}\, \tilde{\boldsymbol{g}}^{(N_\mathrm{L}-1)} \left(  \tilde{\boldsymbol{g}}^{(N_\mathrm{L}-2)} \left(\ \ldots\ \tilde{\boldsymbol{g}}^{(2)} \left( \tilde{\boldsymbol{g}}^{(1)} \left( \boldsymbol{x}^{(0)} \right) \right) \right) \right) +\boldsymbol{\tilde{b}}^{(N_\mathrm{L})}
\end{align}

Independent of the architecture, the PINNs are trained by an Adam optimizer with a constant learning rate of $10^{-4}$ for a prescribed number of $10^6$ steps.

As loss function, a weighted sum of a number of partial losses is defined by

\begin{align}
    \mathscr{L} &= \mathscr{L}_\mathrm{data} + 10^{-1} \mathscr{L}_\mathrm{NS} + 10^{-2} \mathscr{L}_\mathrm{EE} + 10^{-3} \mathscr{L}_\mathrm{CE} + \mathscr{L}_\mathrm{b-p} + 10^{-3} \mathscr{L}_\mathrm{b-s} + 10^{-4} \mathscr{L}_\mathrm{b-T} + 10^{-4} \mathscr{L}_\mathrm{p-c}.
\end{align} 

These partial losses are calculated each training step in the following manners and based on the following sampling rules:
For the data loss $\mathscr{L}_\mathrm{data}$ a random-selected batch of size $N_\mathrm{d}=4096$ is sampled from the provided data (indicated by $\check{\cdot}$) to calculate

\begin{align}
    \mathscr{L}_\mathrm{data} &= \frac{1}{N_\mathrm{d}} \sum_{j=1}^{N_\mathrm{d}} \left| \boldsymbol{u}_j-\check{\boldsymbol{u}}_j  \right|^2.
\end{align} 

Each step, another $N_\mathrm{c}=4096$ sampling locations are random-uniformly distributed throughout the spatio-temporal domain to calculate the loss for the momentum ($\mathscr{L}_\mathrm{NS}$), energy ($\mathscr{L}_\mathrm{EE}$) and continuity ($\mathscr{L}_\mathrm{CE}$) PDEs as well as the pressure-centering loss ($\mathscr{L}_\mathrm{p-c}$). Here, $\nu=\sqrt{\frac{Pr}{Ra}}$ functions as an additional trainable parameter \citep{Mommert2026}, which is trained exclusively by losses originating in $\mathscr{L}_\mathrm{NS}$:

\begin{align}
    \mathscr{L}_\mathrm{NS} &= \frac{1}{N_\mathrm{c}} \sum_{j=1}^{N_\mathrm{c}} \left| \partial_t \boldsymbol{u}_j + (\boldsymbol{u}_j\cdot\nabla)\boldsymbol{u}_j +\nabla p_j - \nu\,\nabla^2\boldsymbol{u}_j - T_j \boldsymbol{e}_z \right|^2 \\
    \mathscr{L}_\mathrm{EE} &= \frac{1}{N_\mathrm{c}} \sum_{j=1}^{N_\mathrm{c}} \left| \partial_t T_j + (\boldsymbol{u_j}\cdot\nabla)T_j - (\nu/\mathrm{Pr})\,\nabla^2 T_j \right|^2 \\
    \mathscr{L}_\mathrm{CE} &= \frac{1}{N_\mathrm{c}} \sum_{j=1}^{N_\mathrm{c}} \left| \nabla \cdot \boldsymbol{u}_j \right|^2\\
    \mathscr{L}_\mathrm{p-c} &= \frac{1}{N_\mathrm{c}} \left| \sum_{j=1}^{N_\mathrm{c}} p_j   \right| \\
\end{align}

Furthermore, boundary points are random uniformly sampled on the walls of the domain for each training step. These points amount to $1024$ for each pair of parallel walls.
Exclusively on the top and bottom surfaces, the thermal boundary conditions are prescribed as the loss $\mathscr{L}_\mathrm{b-T}$ on $N_\mathrm{T}=1024$ points:

\begin{align}
    \mathscr{L}_\mathrm{b-T} &= \frac{1}{N_\mathrm{T}} \sum_{j=1}^{N_\mathrm{T}} \left| T_j-\check{T}_j  \right|^2
\end{align} 

For the velocity boundary conditions, we distinguish between impermeability ($\mathscr{L}_\mathrm{b-p}$) and no-slip ($\mathscr{L}_\mathrm{b-s}$) losses, which apply to all $N_\mathrm{w}=3072$ boundary points. Here, $\boldsymbol{n}$ denotes the wall-normal vector:

\begin{align}
    \mathscr{L}_\mathrm{b-p} &= \frac{1}{N_\mathrm{w}} \sum_{j=1}^{N_\mathrm{w}} \left| \boldsymbol{u}_j\cdot\boldsymbol{n}_j  \right|^2\\
    \mathscr{L}_\mathrm{b-s} &= \frac{1}{N_\mathrm{w}} \sum_{j=1}^{N_\mathrm{w}} \left| \boldsymbol{u}_j-\boldsymbol{u}_j\odot\boldsymbol{n}_j  \right|^2
\end{align}


\subsection{Comparison}\label{sec:comp}

As a benchmark, we test the architecture described in section~\ref{sec:arch} ('\textit{new}') against two conventional, sine-activated MLPs. The first of which has an equal amount of layers and neurons ('\textit{eq\_net}'). This results in approximately half the trainable parameter count of the new architecture, as we count their real and imaginary parts as separate trainable parameters. To take this into account, we also test against a network with a similar parameter count ('\textit{eq\_par}'), which was achieved by multiplying the layer width by a factor of $\sqrt{2}$.
The details of the test cases are gathered in table~\ref{tab:cases}.

\begin{table}[h!]
	\caption{Configuration and metrics of the test cases. (* Referring to an Nvidia RTX\,4090, while the remining cases refer to an Nvidia RTX\,6000\,Ada using the same AD102 die.)}
	\centering
	\begin{tabular}{cccccccc}
		\toprule
		Name     & $N_\mathrm{L}$ & $N_\mathrm{N}$  &  total parameter count $N_\theta$ & wall time per epoch & $\mathrm{MAE}_T$& $\mathrm{PCC}_T$& $\mathrm{R2}_T$ \\
		\midrule
		\textit{eq\_net} & $8$  & $256$ & $397318$ & $\sim0.7\,\mathrm{s}^*$ & $0.0291$ & $0.9351$ & $0.8706$ \\
		\textit{eq\_par} & $8$ & $362$ & $792062$ & $\sim1\,\mathrm{s}$ & $0.0258$ & $0.9518$ & $0.8991$ \\
		\textit{new}     & $8$ & $256$ & $792070$ & $\sim1\,\mathrm{s}$ & $\bold{0.0168}$ & $\bold{0.9775}$ & $\bold{0.9547}$\\
		\bottomrule
	\end{tabular}
	\label{tab:cases}
\end{table}

Table~\ref{tab:cases} further shows the metrics achieved by the test cases.
Regarding the expense in computational time for one epoch, the new architecture performs similar to as sine-activated MLP with the same number of trainable parameters.
Remarkably, it achieves significantly better metrics (lower mean absolute errors (MAE) higher Pearson correlations (PCC) and R2 scores for the reconstruction of the temperature field.

\begin{figure}[p!]
	\centering
    \begin{minipage}[t]{0.245\textwidth}
        \centering
        \colheader{ground truth}
        \includegraphics[trim={0 0 580 0}, clip, width=\linewidth]{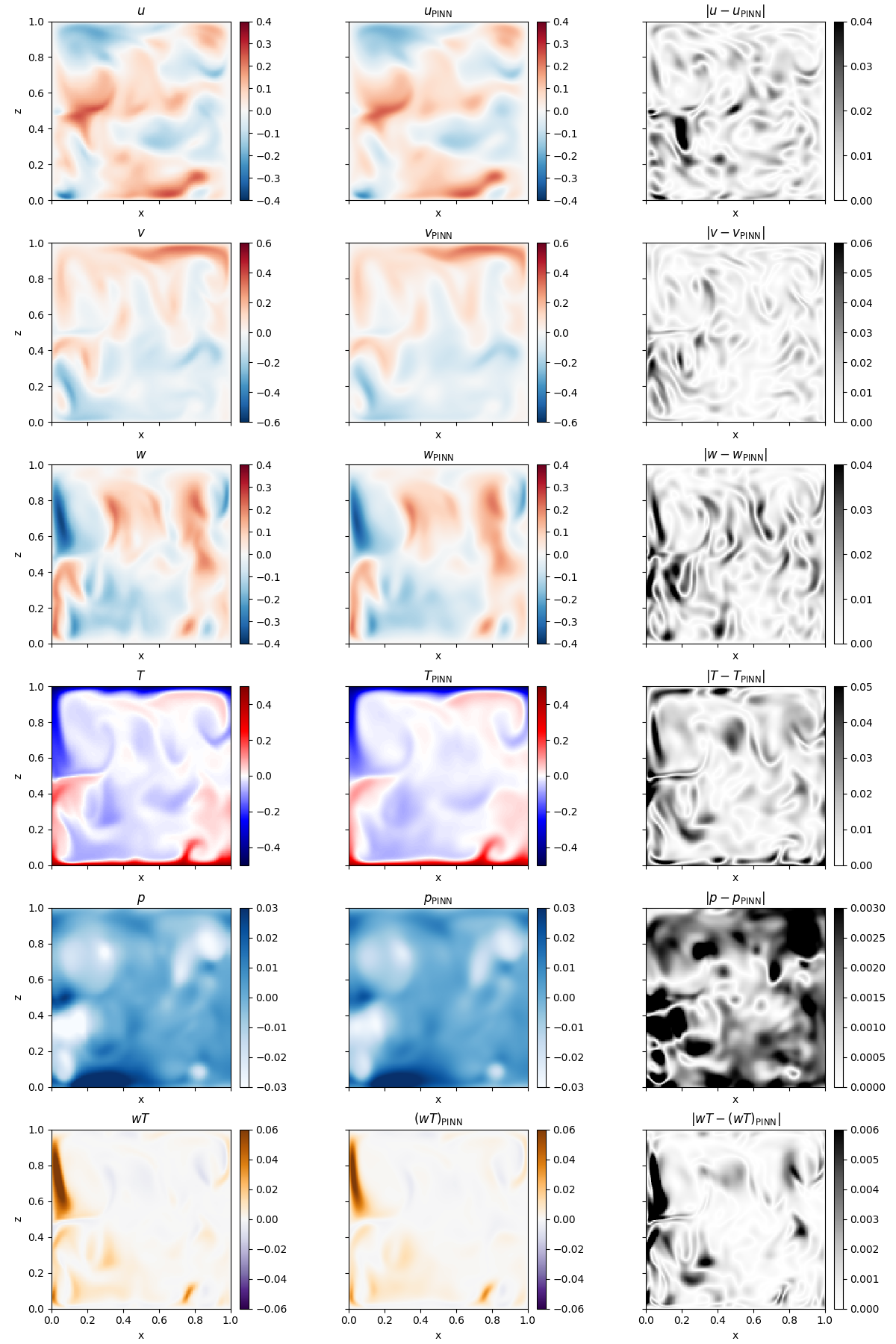}
    \end{minipage}
    \hfill
    \begin{minipage}[t]{0.245\textwidth}
        \centering
        \colheader{new}
        \includegraphics[trim={290 0 290 0}, clip, width=\linewidth]{complex_0_gl_A.png}
    \end{minipage}
    \hfill
    \begin{minipage}[t]{0.245\textwidth}
        \centering
        \colheader{eq\_par}
        \includegraphics[trim={290 0 290 0}, clip,width=\linewidth]{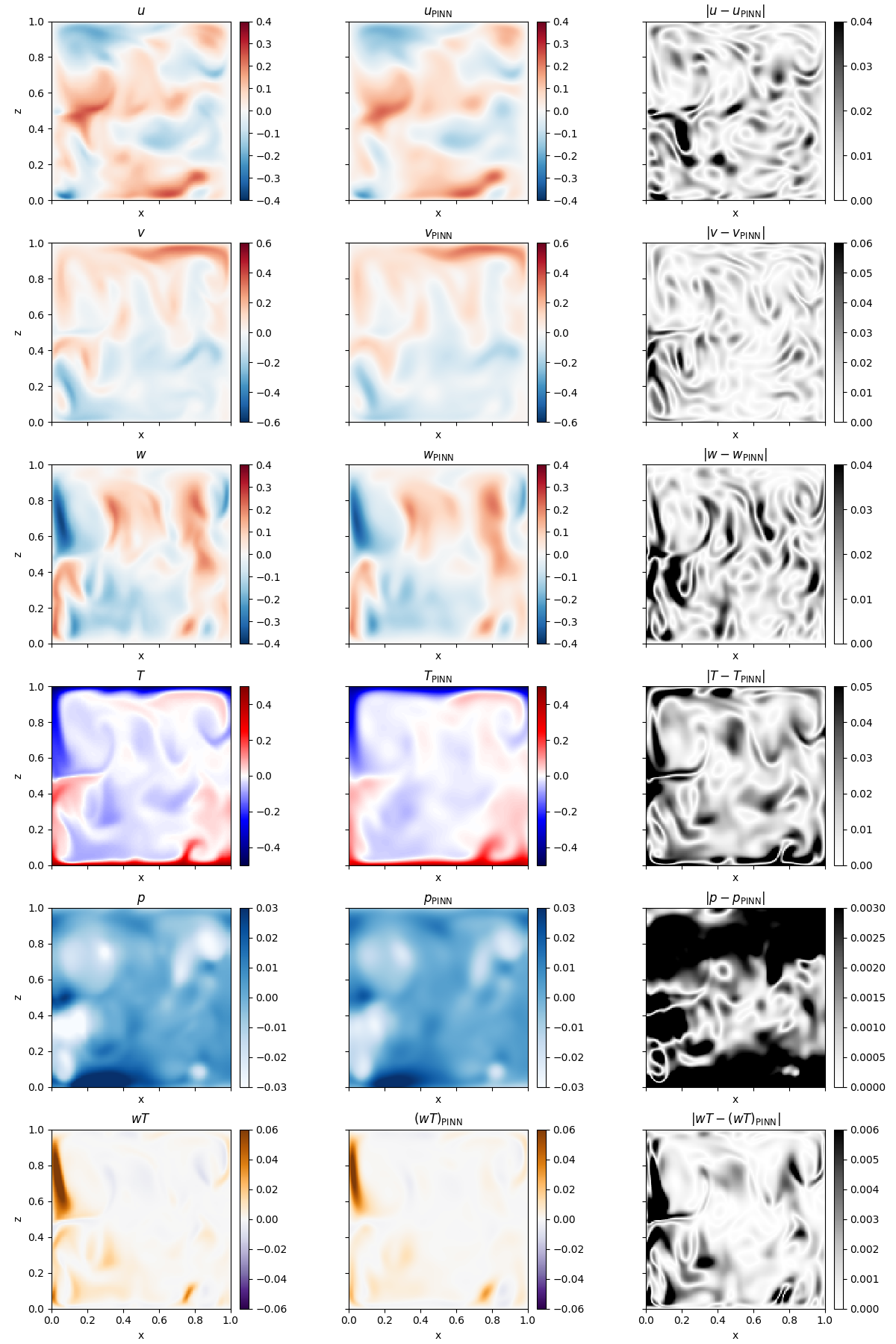}
    \end{minipage}
    \hfill
    \begin{minipage}[t]{0.245\textwidth}
        \centering
        \colheader{eq\_net}
        \includegraphics[trim={290 0 290 0}, clip,width=\linewidth]{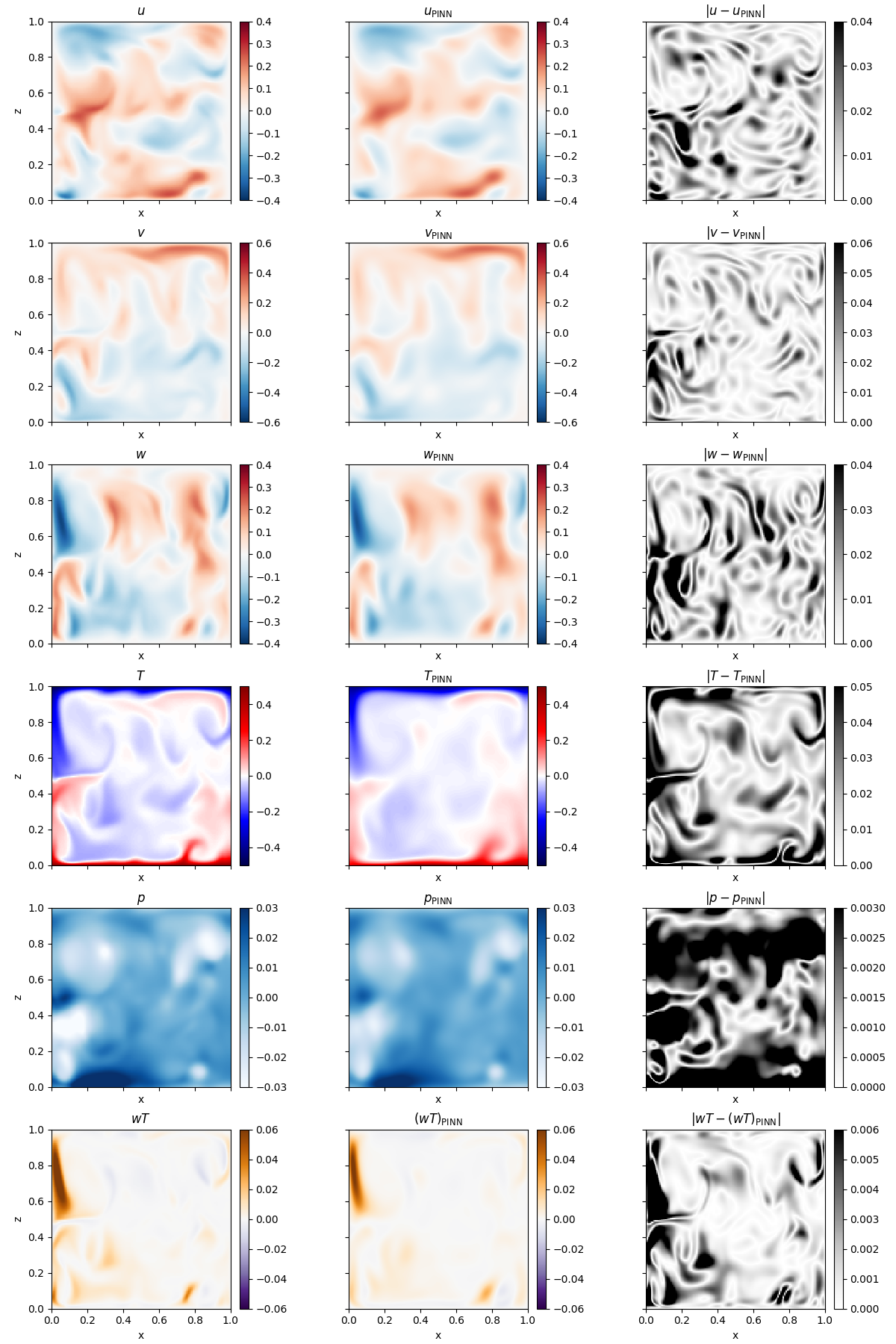}
    \end{minipage}
	\caption{Ground truth and PINN reconstruction for the three test cases in a central vertical section and the central time instance.}
	\label{fig:fields}
\end{figure}

A more detailed view on reconstruction performance is provided by figure~\ref{fig:fields}. It shows the fields of the velocity components ($\boldsymbol{u}=[u\ v\ w ]$), temperature, pressure, and convective heat fluxes ($wT$) for the ground truth and the three test cases in a central vertical section at the central time instance.

Regarding the fields of the velocity components, the PINNs task is to fill the gaps between the sparse data points. This is best achieved by the new architecture. 
While all test cases are able to recover the main velocity structures, small features are rendered best by the new architecture. An example for this can be found in the top right corner of the section for the $w$ component (third row of figure~\ref{fig:fields}), where the pair of an up-welling and a down-welling region looks increasingly blurred for the cases \textit{eq\_par} and \textit{eq\_net}.

This difference in quality also impacts the rendition of the reconstructed fields $T$ and $p$, which show the same trends as for the velocity components. As the task of the complete reconstruction of the fields based on the PDEs is more challenging, the differences between the cases are more pronounced. A prominent example can be found in the bottom right corner of the temperature fields (fourth row of figure~\ref{fig:fields}).
At $x\approx0.8$, a warm plume detaches from the boundary layer. While the new architecture is able to form the head of the mushroom-like structure, this feature is blurred to a degree that the shape of the head changes in the other two cases.
This then culminates in the fields of the convective heat fluxes $wT$ (bottom row of figure~\ref{fig:fields}), for which the the new architecture is able to recover the amplitudes of the various structures the best.

\begin{figure}[h!]
	\centering
    \includegraphics[trim={75 160 680 0}, clip, width=0.5\linewidth]{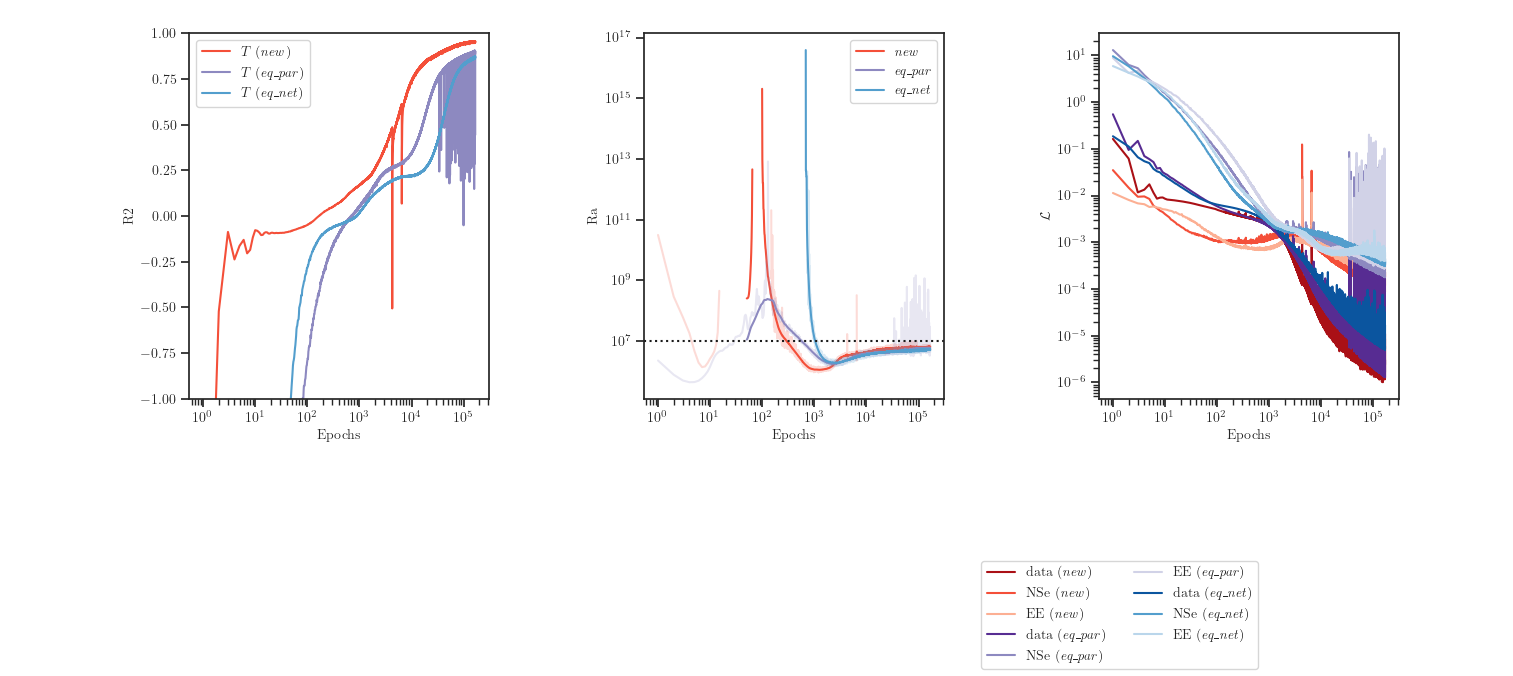}
	\caption{Evolution of the R2 score for the temperature field over the course of the training.}
	\label{fig:train}
\end{figure}

For a better understanding of how these improvements are achieved, figure~\ref{fig:train} shows the evolution of the R2 metric of the temperature over the course of the training of the three test cases.
Besides presenting the final scores noted in table~\ref{tab:cases}, the plot also shows that the new architecture is also better trainable, as significantly less epochs (here equal to $6$ training steps) are required to reach a certain reconstruction quality. 
For the example of $\mathrm{R2}_T=0.75$, the new architecture approximately requires $10^4$ epochs. Compared to that, the sine-activated MLPs require approximately $3\times10^4$ (\textit{eq\_par}) and $6.5\times10^4$ (\textit{eq\_net}) epochs. These render the different epoch-specific wall-times of the different cases insignificant and show that the new architecture provides a faster solution to achieve sufficient temperature reconstruction results. 

Figure~\ref{fig:train} further exhibits instabilities/oscillations of the PINN solution of case \textit{eq\_par}, when its results start  to converge. We observe these oscillations of temperature-reconstruction PINNs in particular for cases with sparse data combined with a high count of trainable parameters. This effect is exaggerated by the logarithmic abscissa, as the PINN solution is typically restored after a few epochs after one of the drop-offs shown in the metric and stays stable for much longer. Nonetheless, the new architecture also appears to be less prone to this effect than the traditional architecture.

\section{Analysis of network structures}
\label{sec:principle}

To understand the reason behind the improvements the new network type exhibits, we consider the expanded forms of a single layer in comparison to one of a simple sine activated network.

A layer of a sine activated MLP (denoted by $\tilde{\cdot}$) with real numbered trainable parameters looks as follows in its extended form

\begin{align}
\tilde{g}^{(L)}_i &= \sin{ \underbrace{ \left( \sum_j \left( \tilde{W}^{(L)}_{ij} \tilde{x}^{(L-1)}_j \right) + \tilde{b}^{(L)}_{i} \right)}_{\tilde{h}^{(L)}_i}}.
\end{align}

We define $\tilde{h}^{(L)}_i$ as the linear part of the transformation imposed by the layer $L$. For the sake of later comparison, we replace the outputs of the previous layer $\tilde{x}^{(L-1)}_j$ by its sine-activated linear part, which yields

\begin{align}
\tilde{g}^{(L)}_i &= \sin{\left( \sum_j \left( \tilde{W}^{(L)}_{ij} \sin{\left(\tilde{h}^{(L-1)}_j\right)}  \right) + \tilde{b}^{(L)}_{i} \right)}.\label{eq:exp_sine}
\end{align}

For the architecture described in section~\ref{sec:arch} with complex-numbered parameters, the equivalent expansion is as follows:

\begin{align}
g^{(L)}_i &= \cos{ \underbrace{ \left( \sum_j \left( \Im\left(W^{(L)}_{ij}\right)  \Re\left( x^{(L-1)}_j\right) + \Re\left(W^{(L)}_{ij}\right)  \Im\left( x^{(L-1)}_j\right) \right) + \Im\left(b^{(L)}_{i}\right)\right)}_{h^{(L)}_i}  } + i \sin{\left(h^{(L)}_i  \right)}\\
g^{(L)}_i &= \cos{ \left( \sum_j \left( \Im\left(W^{(L)}_{ij}\right) \cos{\left(h^{(L-1)}_j\right)} + \Re\left(W^{(L)}_{ij}\right)  \sin{\left(h^{(L-1)}_j\right)} \right) + \Im\left(b^{(L)}_{i}\right)\right)  } + i \sin{\left(h^{(L)}_i  \right)}
\end{align}

The insertion of the cosine and sine activated linear parts $h^{(L-1)}_j$ of the previous layer results in a weighted sum of a sine and a cosine with the same argument inside the sum over the $j$ layer inputs.
This cosine-sine sum can be rewritten as a single sine function, which has an amplitude $\left| W^{(L)}_{ij}\right|$ and phase shift $\varphi^{(L)}_{ij}$,

\begin{align}
g^{(L)}_i &= \cos{ \left( \sum_j\left( \left| W^{(L)}_{ij}\right|  \sin{\left(h^{(L-1)}_j + \underbrace{\mathrm{arctan2}\left(\Im\left(W^{(L)}_{ij}\right),\Re\left(W^{(L)}_{ij}\right)\right)}_{\varphi^{(L)}_{ij}}\right) }  \right) + \Im\left(b^{(L)}_{i}\right)\right)  } + i \sin{\left(h^{(L)}_i  \right)}.
\end{align}

By inserting $\varphi^{(L)}_{ij}$ as symbol for the weight-depended phase shift, we obtain a form similar to that in equation~\ref{eq:exp_sine}:
\begin{align}
g^{(L)}_i &= \cos{ \left( \sum_j\left( \left| W^{(L)}_{ij}\right|  \sin{\left(h^{(L-1)}_j + \varphi^{(L)}_{ij}\right) }  \right) + \Im\left(b^{(L)}_{i}\right)\right)  } + i \sin{\left(h^{(L)}_i  \right)}\label{eq:exp_cexp}
\end{align}

Comparing the equations~\ref{eq:exp_sine} and \ref{eq:exp_cexp} allows us to understand the improvements achieved by implementing a complex exponential activation.
By carrying both a cosine and a sine function in the complex output, they are recombined in the linear combination of the subsequent layer.
This recombination can be expressed in the form $\left|W^{(L)}_{ij}\right| \sin{\left(h^{(L-1)}_j + \varphi^{(L)}_{ij}\right)}$, in which the periodic activation of the layer $L-1$ remarkably carries a phase shift depending on the parameters of the layer $L$.
Regarding the training of these networks, this means that optimizing phase shifts requires less back-projection steps, which might be beneficial.
However, we presume a larger effect caused by the fact, that this arrangement also means that one periodic output of the layer $L-1$ can be used with different phase shifts for each of the $i$ neurons of layer $L$. In contrast, a sine-activated MLP could only provide periodic functions with the same frequencies or wavenumbers and different phase shifts in a "one per neuron of the layer $L-1$" manner, if they were required by layer $L$.
The results of section~\ref{sec:comp} show that PINNs expressing fluid flows seem to benefit from this synergy effect of being able to use single latent periodic features with multiple different phase shifts as outputs.

\section{Conclusion \& Outlook}
We presented an improved architecture for physics-informed neural networks with the task of reconstructing hidden fields of fluid flows. Its key feature is the use of the complex exponential form to express periodic activations of the otherwise standard multi-layer perceptron architecture.
We found this architecture significantly beneficial in comparison with networks using the traditional sine-activations to introduce periodicity for a test case of temperature reconstruction from sparse velocity data in cubic Rayleigh-Bénard convection at a Rayleigh number of $10^7$ and Prandtl number of $0.7$. 
For this particular test case, the mean absolute error was reduced by over $30\,\%$ compared to a sine-activated network with the same number of layers and total parameter count.
Further, the respective $\mathrm{R2}$ score was increased from approximately $0.9$ to $0.95$.
The results also showed that the complex-exponential-based architecture requires only approximately a third of the optimization steps to reach results of similar quality, which directly translates into total wall-time, as the differences in wall time per epoch are low in comparison.
An expansion of the formulation of a single layer showed that these improvements root in the communication of sine-cosine pairs from layer to layer. This allows the adaptation of the phase of the latent periodic function deeper into the forward pass. Another remarkable advantage is that the presented architecture allows one layer the use of single latent periodic functions with different phases for each of its neurons.
Given the results for the reconstruction task, this synergy of periodic functions with different phases appears to fit the representation of turbulent flows very well.
Future research should therefore explore the use of this architecture type for a wider range of applications.

\section*{Acknowledgments}
The authors gratefully acknowledge the scientific support and HPC resources provided by the German Aerospace Center (DLR). The HPC system CARA is partially funded by "Saxon State Ministry for Economic Affairs, Labour, Energy and Climate Action" and "Federal Ministry of Research, Technology and Space".
M.M. gratefully acknowledges the financial support of the German Research Foundation (DFG) under grant MO 5240/1-1.

\appendix

\section{Weight matrix initialization}\label{sec:init}

As stated in section~\ref{sec:arch}, only the imaginary parts of the weight matrices undergo an initialization which samples from a uniform distribution, while their real parts are initialized as zeros.
This appendix aims to provide the data for this decision, in contrast to applying the same initialization strategies to both real and imaginary parts of the weight matrices.
For that, we compare the test case \textit{new} which features the initialization described in section~\ref{sec:arch} against a network which uses the same random initializers for the real parts as for the imaginary parts. To rule out random advantages, this comparison was done for the data set of sampling points used in the main part (A), as well as for a second data set with the same density but different data sampling positions (B).

The results of the key performance indicators are gathered in table~\ref{tab:init}. These results show that the presented network architecture still achieves better results than traditional sine-activated MLPs, even when a suboptimal initialization scheme is used.
However, the use of the better initialization scheme is a main contributing factor to achieve the improvements exhibited in section~\ref{sec:comp}.
Also, the choice of the initialization strategy impacts the results stronger than the variance rooted in different random samplings of the data provided to the PINN.

\begin{table}[h!]
	\caption{Metrics achieved for the architecture based on complex exponential activation for two different initialization strategies of the real part of the weight matrices, namely "zeros" and "same" for to different random samplings of the data of the test case. (* marks the case presented in section~\ref{sec:comp})}
	\centering
	\begin{tabular}{cccccc}
		\toprule
		& initialization & sampling & $\mathrm{MAE}_T$& $\mathrm{PCC}_T$& $\mathrm{R2}_T$ \\
		\midrule
		* & zeros & A  & $0.0168$ & $0.9775$ & $0.9547$ \\
		   & same & A  & $0.0216$ & $0.9732$ & $0.9387$ \\
		   & zeros & B  & $0.0161$ & $0.9842$ & $0.9670$ \\
		   & same & B  & $0.0208$ & $0.9720$ & $0.9420$ \\
		\bottomrule
	\end{tabular}
	\label{tab:init}
\end{table}

\begin{figure}[h!]
	\centering
    \includegraphics[trim={75 160 650 0}, clip, width=0.4\linewidth]{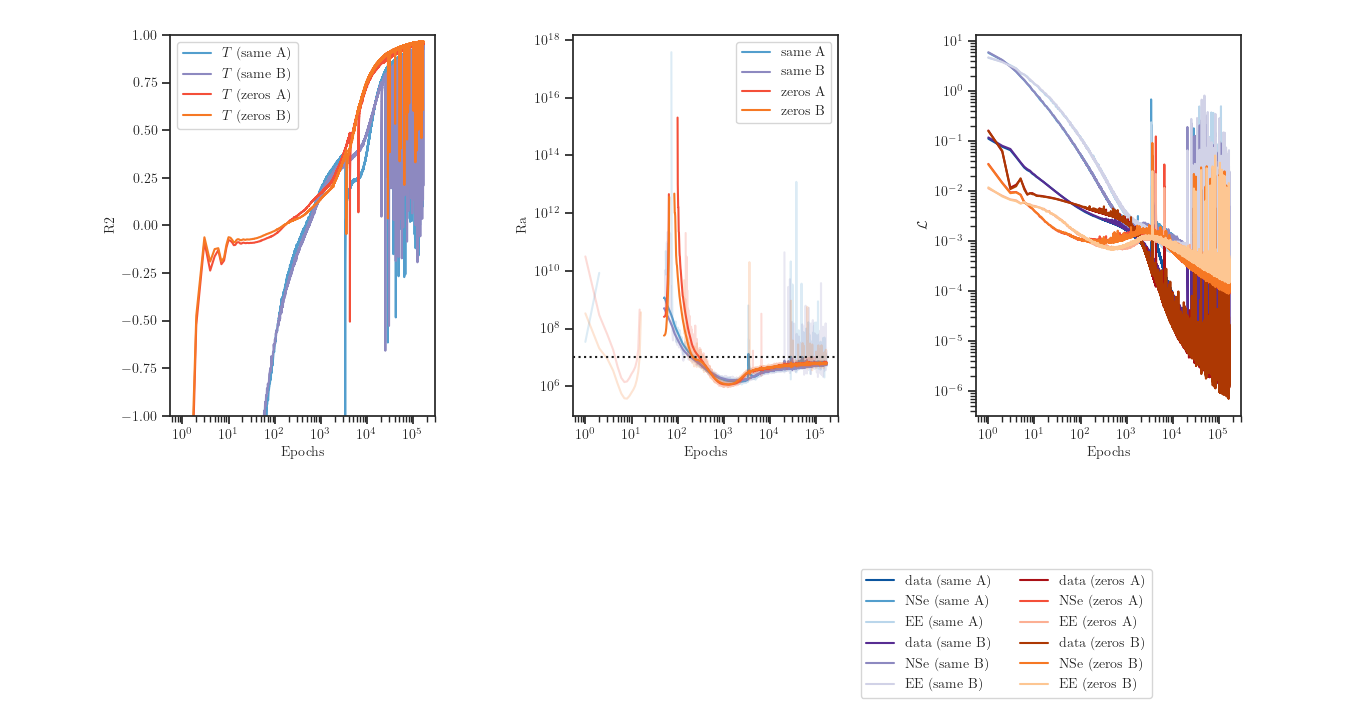}
	\caption{Evolution of the R2 score for the temperature field over the course of the training for different initialization strategies of the real part of the weight matrices and data samplings.}
	\label{fig:init}
\end{figure}

In addition to the metrics at the end of the training, figure~\ref{fig:init} shows the evolution of the $\mathrm{R2}$ score for the four cases discussed above.
It shows that, depending on the sampling, the presented network architecture with the better initialization can also be exposed to the oscillations in the convergence phase. Nonetheless, these solution oscillations are much more pronounced for the suboptimal training strategy, in both the A and B cases.

An interpretation of these is results the following: The pair of real and imaginary parts of the weight matrix elements both have an influence on the amplitude $\left| W^{(L)}_{ij}\right|$ and phase $\varphi^{(L)}_{ij}$ used to combine cosine-sine pairs from the previous layer according to equation~\ref{eq:exp_cexp}. The training of the network is more stable and successful if one part of the cosine-sine pair is dominant during their combination and changes in amplitude and phase are only slowly introduced by the less dominant part.
\newpage

\bibliographystyle{unsrtnat}
\bibliography{ML}






\end{document}